\documentclass[pdflatex,sn-mathphys-num]{sn-jnl}

\usepackage{graphicx}%
\usepackage{multirow}%
\usepackage{amsmath,amssymb,amsfonts}%
\usepackage{amsthm}%
\usepackage{mathrsfs}%
\usepackage[title]{appendix}%
\usepackage{xcolor}%
\usepackage{textcomp}%
\usepackage{manyfoot}%
\usepackage{booktabs}%
\usepackage{algorithm}%
\usepackage{algorithmicx}%
\usepackage{algpseudocode}%
\usepackage{listings}%
\usepackage{longtable}
\usepackage{array}

\theoremstyle{thmstyleone}%
\theoremstyle{thmstyletwo}%

\theoremstyle{thmstylethree}%

\begin{document}

\title[Article Title]{A decision-basis contract for auditable LLM-assisted medical billing verification: deterministic rules, verbatim evidence, and fail-closed abstention}


\author[1]{\fnm{Jan} \sur{H\"olter}}

\author[1]{\fnm{Kevin} \sur{Geis}}

\author[1]{\fnm{Benjamin} \sur{Raab}}
\author*[1]{\fnm{Boris} \sur{Bauke}}\email{boris.bauke@th-ab.de}

\affil[1]{\orgdiv{Competence Center Artificial Intelligence}, \orgname{Aschaffenburg University of Applied Sciences}, \orgaddress{\street{Würzburger Straße 45}, \city{Aschaffenburg}, \postcode{63743}, \state{Bavaria}, \country{Germany}}}




\abstract{This work presents a proof of concept for auditable LLM-assisted medical billing verification based on a decision-basis contract. The contract separates deterministic checks of versioned fee-catalog rules from LLM-based assessment of free-text documentation. The deterministic layer resolves the applicable catalog release and checks code availability, quantity limits, and exclusions. The semantic layer classifies each claimed item as supported, contradicted, or missing required information. Support and contradiction require a verbatim evidence span; unavailable rule context, unsuccessful assessment, or missing required evidence prevents support through fail-closed abstention. We evaluated four locally run open-weight models on a synthetic catalog and 36 curated cases under the contract, an ablation without explicit documentation requirements, and an end-to-end baseline. Outcome agreement varied across models and showed no consistent advantage over the baseline. Explicit documentation requirements improved identification of missing information for all four models. The evidence gate also exposed cases in which correct raw judgments lacked valid evidence and were converted to incomplete decision-basis entries. The results show how explicit decision records can make rule findings, documentation judgments, and abstention reasons inspectable. Evaluation on real catalogs, independently annotated documentation, and with human reviewers is required to assess practical value.}




\maketitle

\section{Introduction}\label{sec:intro}

Billing verification determines whether claimed services are supported by the applicable fee catalog and the case documentation. Formal checks determine whether the catalog release valid on the treatment date contains a claimed code, whether the claimed quantity remains within its maximum, and whether claimed items exclude each other. Semantic checks determine whether the free-text documentation describes the claimed service with the attributes required by the catalog. This distinction motivates separating deterministic rule checks from documentation assessment.

Large language models (LLMs) can assist with documentation assessment. Deploying open-weight models locally allows sensitive treatment documentation to be processed within an institution's infrastructure. However, benchmarks for medical code querying have shown errors in model-generated codes~\cite{soroush2024}. The primary design challenge for verification is therefore making the model's contribution inspectable and to define how unavailable or unacceptable assessments affect the decision. A reviewer must be able to trace an outcome back to its underlying rule findings and documentation judgments, including specific reasons for abstention.

This paper presents a proof of concept for a two-layer decision-based contract. The deterministic layer resolves the applicable catalog release and rules before checking quantity limits and exclusions. The semantic layer assesses each claimed item with an applicable rule as supported, contradicted, or missing required information. Support and contradiction verdicts are retained only when accompanied by a verbatim span from the documentation; missing-information verdicts identify the absent attributes. Rule findings and effective documentation judgments form the decision basis, from which a fixed precedence determines the final outcome. If rule context is unavailable or if technical/evidence failures are detected, the system precludes support and leads to abstention—unless another basis entry already requires rejection. This is the scope of fail-closed behavior in this paper.

The evaluation examines outcome agreement together with the contract's decision basis and processing diagnostics. Across four locally run open-weight models, agreement with the reference was higher under \emph{system} than under the end-to-end \emph{baseline} for some models and lower for others. Diagnostics reveal how evidence requirements and outcome derivation affect the final decision. Auditability is operationalized through these inspectable records.

The contributions of this work are: (i) a decision-basis contract that separates deterministic rules from LLM-based assessment, requires verbatim evidence for support and contradiction, and prevents support when required parts of the decision basis are unavailable (Section~\ref{sec:contract}); (ii) a synthetic catalog and a reference dataset constructed to exercise selected mechanisms of the contract (Section~\ref{sec:catalogdataset}); and (iii) an evaluation of four open-weight models comparing the contract approach against an ablation without explicit documentation requirements and an end-to-end baseline (Sections~\ref{sec:designeval} and~\ref{sec:results}). Section~\ref{sec:relatedwork} reviews related work, Section~\ref{sec:discussion} discusses the findings and limitations, and Section~\ref{sec:conclusion} concludes the paper.

\section{Related work}\label{sec:relatedwork}
Rule-based computer-assisted coding and audit systems are established in practice. They ensure formal consistency but resolve semantic conflicts between documentation and claimed codes only to a limited extent. Generic LLMs used as stand-alone coders frequently produce imprecise or fabricated codes~\cite{soroush2024}. In a Scandinavian randomized crossover trial of an assistive coding tool in workflow, median coding time dropped by 46\% for longer texts, but accuracy did not improve significantly~\cite{chomutare2025}. Because model performance still lags behind human coders, targeted use has been recommended: code suggestion, audit support, and the deliberate assignment of well-delimited case groups~\cite{gan2025}. For the delimited task of assigning German GO\"A codes to radiology reports, fine-tuned models reached competitive multi-label performance, while manual review was still recommended~\cite{arzideh2026}.
The field has been shifting toward verification-oriented, neuro-symbolic, and human-supervised procedures. Verification steps reduce hierarchically close but billing-relevant misassignments~\cite{yuan2025}, and combining an LLM with a rule-based expert system can establish an auditable decision chain~\cite{prenosil2025}. CPTCoder pairs constrained decoding with conflict rules and interpretable confidence scores~\cite{wang2026}. GuideTree uses evidence-bound review trees that are versioned and regression-tested over time~\cite{ge2026}.
Retrieval-augmented generation is another direction being explored alongside these verification approaches, aiming to ground LLM outputs in external medical knowledge. Evaluation practices for this approach remain insufficiently standardized~\cite{amugongo2025}. 
Conceptually related approaches also comes from outside medical coding - particularly claim verification and secure system design. FEVER operationalizes claim verification by labeling claims as Supported, Refuted, or NotEnoughInfo~\cite{thorne2018}. In security engineering, RFC~4949 defines fail-secure behavior defines fail-secure behavior as the preservation of a secure state when a failure is detected or occurs~\cite{shirey2007}. This concept has recently been adapted for LLM alignment as "fail-closed" alignment~\cite{coalson2026}.

\section{Methods}\label{sec:methods}

\subsection{Problem and decision contract}\label{sec:contract}

A billing case $C$ includes an effective date, free-text documentation $D$, and a list of claimed items $I$ (each containing a code and count). A rule catalog $K$ consists of releases. Each release has a validity interval $[\mathit{valid\_from}, \mathit{valid\_to})$ and contains entries indexed by code. Each entry specifies a legend, a maximum count, a symmetric exclusion list, and the required documentation attributes. The goal is to decide — for the case as a whole — whether the rules and the provided documentation support every claim in $I$, and to produce an inspectable record of \emph{why} that decision was reached. The outcome falls into one of four categories: \texttt{SUPPORTED}, \texttt{NOT\_SUPPORTED} (at least one contradiction or rule violation), \texttt{INSUFFICIENT\_DOCUMENTATION} (no contradiction, rule violation, or unavailability, but a required attribute is not documented), and \texttt{NOT\_EVALUABLE} (the rule basis or the assessment is not reliably available).

The contract separates the decision into two layers according to the input required by each layer. The deterministic layer answers every question that depends only on structured data: which release applies to the date, whether each code has an entry, whether the summed count per code exceeds its maximum, and whether two claimed codes exclude each other. The semantic layer answers exactly one question for each item with an applicable rule: \emph{Does the documentation support this item under this rule?} Answering this requires understanding language, so the task is delegated to an assessor working through a fixed interface. In this implementation, that assessor is a LLM. Every result from either layer is recorded as a basis entry with a basis code, the affected items, structured facts, and any retained verbatim evidence spans. Table~\ref{tab:basis_codes} lists the basis codes, the layer that produces each code, and the outcome each code implies.

\begin{table}[h]
\caption{Basis codes of the decision contract with the abbreviation used in Algorithm~\ref{alg:core}, the layer that produces them, and the outcome each implies}\label{tab:basis_codes}
\begin{tabular*}{\textwidth}{@{\extracolsep\fill}llll}
\toprule
Basis code & Abbr. & Layer & Outcome \\
\midrule
\texttt{RULE\_CHECKS\_PASSED}          & \texttt{RULES\_OK}   & rule      & \texttt{SUPPORTED}                   \\
\texttt{MAX\_COUNT\_EXCEEDED}          & \texttt{MAX\_COUNT}  & rule      & \texttt{NOT\_SUPPORTED}              \\
\texttt{EXCLUSION\_CONFLICT}           & \texttt{EXCLUSION}   & rule      & \texttt{NOT\_SUPPORTED}              \\
\texttt{NO\_APPLICABLE\_RULE\_CONTEXT} & \texttt{NO\_RULE}    & rule      & \texttt{NOT\_EVALUABLE}              \\
\texttt{RULE\_VERSION\_UNRESOLVED}     & \texttt{NO\_RELEASE} & rule      & \texttt{NOT\_EVALUABLE}              \\
\texttt{DOCUMENTATION\_SUPPORT}        & \texttt{SUPPORT}     & semantic  & \texttt{SUPPORTED}                   \\
\texttt{DOCUMENTATION\_CONTRADICTION}  & \texttt{CONTRA}      & semantic  & \texttt{NOT\_SUPPORTED}              \\
\texttt{REQUIRED\_ATTRIBUTE\_MISSING}  & \texttt{MISSING}     & semantic  & \texttt{INSUFFICIENT\_DOCUMENTATION} \\
\texttt{DECISION\_BASIS\_INCOMPLETE}   & \texttt{INCOMPLETE}  & system only & \texttt{NOT\_EVALUABLE}              \\
\botrule
\end{tabular*}
\end{table}

To perform the semantic judgment, the LLM is provided only with necessary data: the case documentation, the item's code and count (including the total claimed), and the applicable rule’s legend and requirements. Identifiers such as the item identifier or catalog version are withheld from the model. It returns exactly one of three verdicts, \texttt{DOCUMENTATION\_SUPPORT}, \texttt{DOCUMENTATION\_CONTRADICTION}, or \texttt{REQUIRED\_ATTRIBUTE\_MISSING}, together with an explanation, a list of evidence spans, and a list of missing attributes. If a valid assessment cannot be reached, the assessor adapter signals unavailability and provides a diagnostic category. The basis code \texttt{DECISION\_BASIS\_INCOMPLETE} is reserved for the system logic and cannot be returned by the assessor.

The Algorithm~\ref{alg:core} evaluates a single case, assuming that the assessor complies with the specified interface and that both the case and catalog inputs are structurally validated. The system interprets an absent upper validity bound as $+\infty$. Release resolution requires exactly one valid release. Otherwise, the case is classified as \texttt{NOT\_EVALUABLE}, and the assessor is not invoked (lines~\ref{ln:release}--\ref{ln:noassessor}). The rule layer (lines~\ref{ln:rules_begin}--\ref{ln:rules_end}) records any rule-context issue or violation; if none are found, it records \texttt{RULE\_CHECKS\_PASSED}.
The semantic layer (lines~\ref{ln:semantic_begin}--\ref{ln:semantic_end}) calls the assessor once for each item with an applicable rule. Two mechanisms make this layer fail-closed. First, an unavailable assessment is converted to \texttt{DECISION\_BASIS\_INCOMPLETE}, with the diagnostic category recorded as its cause (line~\ref{ln:unavailable}). Second, the evidence gate (lines~\ref{ln:gate}--\ref{ln:gate_end}) retains only spans that occur verbatim in the documentation and contain at least one alphanumeric character. Any \texttt{SUPPORT} or \texttt{CONTRA} verdict lacking a retained span is downgraded to \texttt{DECISION\_BASIS\_INCOMPLETE} with the cause \texttt{NO\_EVIDENCE} (\texttt{MISSING\_VALID\_EVIDENCE\_SPAN} in the implementation). To ensure model behavior and evidence-gate effects remain distinguishable during evaluation, the system retains the raw verdict alongside the effective basis code.

\begin{algorithm}[t]
\caption{Decision-basis evaluation of one billing case. Basis codes are
abbreviated as in Table~\ref{tab:basis_codes}.}\label{alg:core}
\footnotesize
\begin{algorithmic}[1]
\raggedright
\Require case $(\mathit{date}, D, I)$, catalog $K$, assessor $A$
\State $B \gets \emptyset$
\State $R \gets \{\, r \in K : r.\mathit{from} \le \mathit{date} < r.\mathit{to} \,\}$ \label{ln:release}
\If{$|R| \ne 1$} add \texttt{NO\_RELEASE} to $B$; \Return $\Call{Outcome}{B}, B$ \label{ln:noassessor}
\EndIf
\State $E \gets$ entries of the single release in $R$, by code
\Statex \textit{Deterministic rule layer}
\State $\mathit{total}[c] \gets \sum_{i \in I,\, i.\mathit{code}=c} i.\mathit{count}$ \label{ln:rules_begin}
\ForAll{codes $c$ in $\mathit{total}$}
    \If{$c \notin E$} add \texttt{NO\_RULE}$(c)$ to $B$ \label{ln:unknown}
    \ElsIf{$\mathit{total}[c] > E[c].\mathit{max}$} add \texttt{MAX\_COUNT}$(c)$ to $B$
    \EndIf
\EndFor
\ForAll{unordered pairs $\{c_1, c_2\}$ of claimed known codes with $c_2 \in E[c_1].\mathit{excl}$}
    add \texttt{EXCLUSION}$(c_1, c_2)$ to $B$
\EndFor
\If{$B = \emptyset$} add \texttt{RULES\_OK}$(I)$ to $B$ \label{ln:rules_end}
\EndIf
\Statex \textit{Semantic layer: one assessor call per item with a rule}
\ForAll{$i \in I$ with $i.\mathit{code} \in E$} \label{ln:semantic_begin}
    \State $a \gets A(D,\; i.\mathit{code},\, i.\mathit{count},\, \mathit{total}[i.\mathit{code}],\, E[i.\mathit{code}].\{\mathit{code}, \mathit{legend}, \mathit{reqs}\})$
    \If{$A$ signalled \textsc{Unavailable}$(\mathit{cat})$}
        add \texttt{INCOMPLETE}$(i, \mathit{cat})$ to $B$; \textbf{continue} \label{ln:unavailable}
    \EndIf
    \State $S \gets \{\, s \in a.\mathit{spans} : s \text{ occurs verbatim in } D \text{ and contains an alphanumeric character} \,\}$
    \Comment{evidence gate} \label{ln:gate}

    \If{$a.\mathit{code} \in \{\texttt{SUPPORT}, \texttt{CONTRA}\}$ \textbf{and} $S = \emptyset$}
        add \texttt{INCOMPLETE}$(i, \texttt{NO\_EVIDENCE})$ to $B$ \label{ln:gate_end}
    \Else\ add $a.\mathit{code}(i, S, a.\mathit{missing})$ to $B$
    \EndIf
\EndFor \label{ln:semantic_end}
\State \Return $\Call{Outcome}{B}, B$
\Statex
\Function{Outcome}{$B$} \label{ln:map}
    \State $O \gets \{\, \mathrm{outcome}(b.\mathit{code}) : b \in B \,\}$
        \Comment{Table~\ref{tab:basis_codes}}
    \State \Return first of \texttt{NOT\_SUPPORTED}, \texttt{NOT\_EVALUABLE},
    \texttt{INSUFFICIENT\_DOCUMENTATION} in $O$, else \texttt{SUPPORTED} \label{ln:precedence_end}
\EndFunction
\end{algorithmic}
\end{algorithm}

The final outcome is determined by selecting the most severe result among all basis entries, following the precedence: \texttt{NOT\_SUPPORTED}, \texttt{NOT\_EVALUABLE}, \texttt{INSUFFICIENT\_DOCUMENTATION} (lines ~\ref{ln:map}–~\ref{ln:precedence_end}).
\texttt{SUPPORTED} is only reachable if every claimed item carries \texttt{DOCUMENTATION\_SUPPORT} and if the rule checks passed. Algorithm~\ref{alg:core} guarantees this completeness by construction for valid inputs and contract-conforming assessments. The implementation additionally enforces it as an invariant. This ensures that the deterministic layer remains independent of the assessor and that any non-\texttt{SUPPORTED} outcome is traceable to explicit basis entries. It also prevents signaled assessment failures from silently resulting in a \texttt{SUPPORTED} verdict.

\subsection{Synthetic catalog and reference dataset}\label{sec:catalogdataset}
The rule catalog is fully synthetic and modeled on the structure of the German GO\"A fee schedule, though it does not serve as a legal basis for billing. Each release contains 14 codes in five groups: consultation; examination; blood sampling, injection and infusion; wound care; and sonography. Entries specify a legend, documentation requirements, a maximum count, and symmetric exclusions. The releases cover the intervals cover $[2024\text{-}01\text{-}01,2025\text{-}01\text{-}01)$ and $[2025\text{-}01\text{-}01,+\infty)$. Their rule content differs only in the maximum count of \texttt{I01} (venous blood sampling), which increases from two to three. This catalog is reproduced in Appendix~\ref{secB1} and provided to the \emph{baseline} described in Section~\ref{sec:designeval}.

The dataset contains 36 cases with 83 claimed items (two or three items per case). Each case includes the expected outcome, the applicable release or its absence, and rule basis codes. Items assessed under an applicable rule additionally have a reference verdict and a short rationale: 55 support, 14 required-attribute-missing, and 10 contradiction labels (79 in total). Four items lack a semantic reference label because they either contain unknown codes in otherwise evaluable cases or belong to a case dated before the first release. The algorithm does not invoke the assessor for these items.

The cases and reference verdicts were drafted with Claude Fable 5
(Anthropic), using the catalog and decision contract, then jointly curated by two authors. The resulting evaluation is organized into four case groups based on the intended source of the decision (Table~\ref{tab:reference_axes}).

\begin{table}[h]
\caption{Reference case groups and expected outcomes. S: supported; NS: not supported; ID: insufficient documentation; NE: not evaluable}\label{tab:reference_axes}
\begin{tabular*}{\textwidth}{@{\extracolsep\fill}lrp{0.52\textwidth}l}
\toprule
Axis & Cases & Defining reference behaviour & Outcomes \\
\midrule
a & 9  & All items supported and rule checks passed & S \\
b & 8  & Rule violations or unavailable rule context; all labelled items supported & 5 NS, 3 NE \\
c & 9  & At least one documentation contradiction & NS \\
d & 10 & Required information missing, with no contradiction or rule-context issue or violation & ID \\
\botrule
\end{tabular*}
\end{table}
The dataset serves as a functional acceptance suite for selected behaviors
of the decision contract. A scenario is represented if a case instantiates
its input conditions and the reference judgments specify the corresponding
expected behavior. The suite covers all five rule-layer basis codes,
quantities at and above a maximum, unknown codes, exclusion conflicts,
simultaneous rule violations, and the release-dependent limit of
\texttt{I01}. Temporal scenarios include a date before the first release,
the last valid day of the first release, and the first valid day of the
second release. Case b06 requires summing quantities across items to
detect a maximum violation; c08 combines a contradiction with missing
information and expects the former to determine the outcome; c09 satisfies
the catalog maximum but claims a total above the quantity explicitly
documented.
Appendix~\ref{secB1} contains all 36 cases, including verbatim documentation, effective dates, claims, and scoring reference values. Technical assessment failures and evidence-gate downgrades are recorded during evaluation as described in Section~\ref{sec:designeval}.

\subsection{Experimental design and evaluation}\label{sec:designeval}

Every case is evaluated under three conditions that differ in the information supplied to the model and in the allocation of decision tasks. Under \emph{system}, Algorithm~\ref{alg:core} uses the model as assessor, with one assessment per item with an applicable rule. Under \emph{system-ablation}, only the catalog's documentation-requirement fields are removed; the algorithm, assessor prompt, and remaining catalog fields are retained. Under \emph{baseline}, the model receives the effective date, documentation, all claimed items with their counts, and the complete catalog with both releases in a single request. Its prompt supplies the same three semantic verdict definitions and specifies release validity, code applicability, summed-count maxima, exclusions, and outcome precedence. Item verdicts may additionally be \texttt{NO\_APPLICABLE\_RULE\_CONTEXT}. The baseline performs these tasks without rule checks or evidence gate; its case outcome is accepted after response validation and is not recomputed from its item verdicts. All conditions use the same reference labels.

Four models were run locally through Ollama 0.31.1: \texttt{gemma4:31b}, \texttt{medgemma:27b}, \texttt{qwen3:14b}, and \texttt{qwen3:4b}. Requests were configured with a temperature of 0 and a context window of 8192 tokens; other model options, including thinking, were not explicitly set. Transport or response-validation failures were retried at most once, with a request timeout of 300\,s. A \emph{baseline} request that remained unsuccessful was recorded as \texttt{NOT\_EVALUABLE}, with a diagnostic category and \texttt{DECISION\_BASIS\_INCOMPLETE} for every item. Under the system conditions, an unavailable assessment affects its item, and applies the outcome precedence of Section~\ref{sec:contract}. The reported evaluation comprised one run per model over all 36 cases, with conditions executed in the order \emph{system}, \emph{system-ablation}, and \emph{baseline} within each case: 79 item assessments under each system condition and 36 \emph{baseline} assessments per model, excluding retries. No variance across repeated runs was estimated.

Each request was supplied with a JSON schema, and the adapter validated the returned response. The assessor prompt requested evidence spans, missing attributes, an explanation, and a verdict, in this order. The \emph{baseline} prompt requested one indexed explanation and verdict per item, followed by a case explanation and outcome. The assessor and \emph{baseline} system prompts and an example assessor input are given in Appendix~\ref{secA1}.

Expected and produced outcomes were compared using the stored records. For each model and condition, we report exact outcome matches across all 36 cases and false acceptances, defined as a produced \texttt{SUPPORTED} when any other outcome is expected. Cases affected by unsuccessful assessments remain in these counts.
The documentation-requirement comparison uses the same 14 items labeled \texttt{REQUIRED\_ATTRIBUTE\_MISSING} within the 79 reference-labeled items. For each model under \emph{system} and \emph{system-ablation}, we count how many of these items received the correct raw verdict. Raw denotes the validated model verdict before the evidence gate. An unsuccessful assessment of one of these items has no raw verdict; the item remains in the denominator of 14 and is not counted as correct.

For the contract diagnostics, we count items changed by the evidence gate to \texttt{DECISION\_BASIS\_INCOMPLETE}, with the recorded cause
\texttt{MISSING\_VALID\_EVIDENCE\_SPAN}, and how many of their raw verdicts matched the reference. These counts are summed across models under \emph{system} and \emph{system-ablation}. Technical failures are item records without a valid raw assessment, recorded as \texttt{TRANSPORT\_TIMEOUT} or \texttt{ASSESSOR\_ERROR}. They are counted separately across all models and conditions. These counts refer to affected item records, not individual request attempts. Examples selected from the stored records illustrate outcome derivation and evidence-gate handling. Explanations, missing-attribute lists, and the semantic relevance of evidence are not scored separately. No inferential statistics were applied.

\section{Results}\label{sec:results}

Result records were available for all 36 cases under each model and condition. Table~\ref{tab:focused_outcomes} reports outcome agreement and false acceptances. The \emph{system} agreed with the reference on more cases than the \emph{baseline} for \texttt{medgemma:27b} and \texttt{qwen3:14b}, and on fewer cases for \texttt{gemma4:31b} and \texttt{qwen3:4b}. For \texttt{medgemma:27b}, agreement increased from 19 to 22 cases, and false acceptances fell from 14 to 5. Under \emph{system}, no false acceptances were observed for \texttt{gemma4:31b} or \texttt{qwen3:14b} in this run. No condition achieved the highest agreement for every model.

\begin{table}[h]
\caption{Case outcomes. Correct: number of cases whose outcome matches the
reference; false acceptances: number of \texttt{SUPPORTED} outputs when another
outcome was expected. All recorded cases, including outcomes affected by
unsuccessful assessments, are included}
\label{tab:focused_outcomes}
\begin{tabular*}{\textwidth}{@{\extracolsep\fill}llrr}
\toprule
Model & Condition & Correct /36 & False acceptances \\
\midrule
gemma4:31b   & system          & 28 & 0  \\
             & system-ablation & 31 & 2  \\
             & baseline        & 35 & 0  \\
\midrule
medgemma:27b & system          & 22 & 5  \\
             & system-ablation & 20 & 10 \\
             & baseline        & 19 & 14 \\
\midrule
qwen3:14b    & system          & 33 & 0  \\
             & system-ablation & 31 & 3  \\
             & baseline        & 32 & 2  \\
\midrule
qwen3:4b     & system          & 28 & 1  \\
             & system-ablation & 32 & 1  \\
             & baseline        & 32 & 0  \\
\botrule
\end{tabular*}
\end{table}

Removing the documentation-requirement fields reduced agreement on the 14 items labeled \texttt{REQUIRED\_ATTRIBUTE\_MISSING}. Correct raw verdicts fell from 13 to 7 for \texttt{gemma4:31b}, from 5 to 0 for \texttt{medgemma:27b}, from 14 to 11 for \texttt{qwen3:14b}, and from 12 to 11 for \texttt{qwen3:4b}. These counts use the same reference items before the evidence gate, with unsuccessful assessments retained in the denominator. Agreement was higher with explicit requirements on this subset, but not on case outcomes for every model: \texttt{gemma4:31b} and \texttt{qwen3:4b} had higher overall agreement under \emph{system-ablation}.

Case b02 contrasts the \emph{baseline}'s reported case outcome with the outcome derived by the \emph{system}. It claims \texttt{I1}, which is absent from the catalog, alongside a documented wound treatment. The \texttt{qwen3:14b} \emph{baseline} assigned \texttt{NO\_APPLICABLE\_RULE\_CONTEXT} to \texttt{I1} but nevertheless returned \texttt{SUPPORTED} for the case. The \emph{system} recorded the missing rule context in the decision basis and returned the expected \texttt{NOT\_EVALUABLE}. The wound-treatment item was supported in both conditions.

Across all models under \emph{system} and \emph{system-ablation}, the evidence gate changed 21 item verdicts to \texttt{DECISION\_BASIS\_INCOMPLETE}; 19 of these had matched the reference before the gate. Four additional assessments in the run remained unsuccessful after the configured retry, with their causes recorded separately from evidence-gate interventions. In case c01, the documentation explicitly states: ``Eine symptombezogene Untersuchung wurde ausdrücklich nicht durchgeführt.'' Under \emph{system}, \texttt{gemma4:31b} correctly returned \texttt{DOCUMENTATION\_CONTRADICTION} for the claimed examination but supplied no evidence span; its raw response also contained a non-empty list of missing attributes. The gate recorded \texttt{MISSING\_VALID\_EVIDENCE\_SPAN}, and the case became \texttt{NOT\_EVALUABLE} instead of the expected \texttt{NOT\_SUPPORTED}. Under \emph{system-ablation}, the model returned the same raw label with the quoted sentence as evidence, and the case outcome matched the reference. This example demonstrates how evidence provision can change the final outcome even when the raw documentation label remains the same.

\section{Discussion}\label{sec:discussion}
The evaluation shows that the decision contract makes the relationship between rule findings, documentation judgments, and case outcomes inspectable. Case b02 illustrates the consistency enforced by deriving the outcome from the decision basis. While Case c01 shows how the diagnostics identify missing evidence despite a correct raw judgment. These records ensure that any decision made by the contract is traceable. Outcome agreement improved for some models and declined for others, so the findings support the value of an explicit decision basis without establishing a general accuracy advantage. Benefits to human reviewers remain to be evaluated.

This approach follows the trend toward hybrid information extraction and constrained coding systems~\cite{prenosil2025,wang2026}. By specifying which checks are deterministic and which are delegated to the model, and defining how gaps in that information affect the outcome. This provides a common basis for examining both successful and unsuccessful decisions across the evaluated models.

The \emph{baseline} comparison used a compact rule corpus of 28 entries across two releases. It remains an empirical question how the \emph{baseline} would perform with a much larger corpus, or whether retrieval would be necessary. The present comparison changes task allocation, context, and response protocol together, so the results reflect differences between complete approaches rather than isolated components. The contract assumes an available, correctly structured, and versioned catalog. Constructing that catalog from a real rule corpus is a separate task.

Explicit documentation requirements helped all four models identify items with missing required information, although this benefit did not consistently extend to case outcomes. This suggests that the wording of requirements is critical. Short lists such as ``Wundart (klein, oberfl\"achlich, unkompliziert)'' leave room for different readings, so disagreement with a reference label need not reflect a model error alone. Clarifying whether attributes are jointly required or merely illustrative would improve both the assessment task and the reference judgments.

The evidence gate introduces a checkable acceptance criterion, though at the cost of rejecting correct judgments that lack verbatim evidence spans. Its check establishes literal presence rather than semantic relevance: Even a single matching letter or digit can pass. Fail-closed behavior therefore means withholding support when required rule context, a valid assessment, or required evidence is unavailable; it does not ensure the correctness of assessments that pass the checks. Recorded gate interventions show where the evidence protocol, rather than the raw label alone, determined the outcome.

The dataset supports a functional evaluation of selected contract mechanisms. Its short German cases were drafted with an LLM and jointly curated by two authors, without independent annotation. The dataset forms a consensus acceptance suite rather than a sample of billing practice. Furthermore, model comparisons describe single runs at temperature zero with default thinking settings and do not establish a general ranking. Testing on independently annotated practice data and larger rule corpora, together with a study of reviewer use, is required to see how this traceability translates into practical audit support.

\section{Conclusion}\label{sec:conclusion}

The decision-basis contract enables inspectable LLM-assisted billing verification by separating deterministic rules from semantic judgments and deriving outcomes from their recorded basis. By requiring verbatim evidence for support/contradiction and recording assessment failures, the system ensures that decisions are traceable. In the synthetic acceptance suite, outcome agreement showed no consistent advantage over the \emph{baseline}, but explicit documentation requirements improved the identification of missing information across all models. The diagnostics exposed the cost of rejecting correct raw judgments without valid evidence, alongside the consistency enforced by outcome derivation. These findings establish a proof of concept for auditability through explicit decision records. Further evaluation with real catalogs and human reviewers is needed to determine practical utility.

\section*{Declarations}

\subsection*{Conflict of interest/Competing interests}
The authors declare no competing interests.

\subsection*{Funding}
This research did not receive any specific grant from funding agencies in the public, commercial, or not-for-profit sectors.

\subsection*{Generative AI and AI-assisted technologies in the writing process}
During the preparation of this work the author(s) used “DeepL Write”, “Claude 5 Fable” and “GPT-5.6 Sol” in order to concretize statements and align the structure of the argument. Additionally, AI coding assistance was utilized, including for the generation of the cases dataset as described in the text. After using this tool/service, the author(s) reviewed and edited the content as needed and take(s) full responsibility for the content of the published article.

\begin{appendices}

\section{System prompts and schema}\label{secA1}

\begin{lstlisting}[
  caption={Assessor: system prompt},
  label={lst:assessor-system-prompt},
  basicstyle=\small\ttfamily,
  breaklines=true,
  breakatwhitespace=true,
  columns=fullflexible,
  keepspaces=true,
  showstringspaces=false,
  literate=
    {ä}{{\"a}}1
    {ö}{{\"o}}1
    {ü}{{\"u}}1
    {Ä}{{\"A}}1
    {Ö}{{\"O}}1
    {Ü}{{\"U}}1
    {ß}{{\ss}}1
]
Du bewertest ausschließlich, ob eine Falldokumentation die einzelne beanspruchte
Position gemäß der angegebenen anwendbaren Regel semantisch stützt.
claimed_code_total ist die Summe der Mengen aller beanspruchten Positionen
mit dem Code von claimed_item. Nutze sie für die Dokumentationsprüfung.

Antworte ausschließlich im vorgegebenen JSON-Schema und wähle genau ein
basis_code:
- DOCUMENTATION_SUPPORT nur, wenn die Dokumentation die Position positiv belegt.
- DOCUMENTATION_CONTRADICTION nur bei einem dokumentierten Widerspruch oder einer
  ausdrücklich unvereinbaren Alternative.
- REQUIRED_ATTRIBUTE_MISSING, wenn eine für die Zuordnung zur Regel nötige
  Angabe fehlt.
Beurteile die Deckung der gesamten beanspruchten Menge je Code durch die
Dokumentation, auch wenn diese Menge auf mehrere Positionen aufgeteilt ist.
Ein dokumentierter Leistungsvorgang darf nicht mehrfach zur Deckung derselben
Codebeanspruchung gezählt werden.
Das bloße Fehlen eines Widerspruchs ist keine positive Evidenz.
Widersprechen sich Angaben in der Dokumentation zu einer geforderten Angabe,
gilt sie als nicht belastbar dokumentiert: Wähle REQUIRED_ATTRIBUTE_MISSING
und nenne die Angabe.

Für DOCUMENTATION_SUPPORT und DOCUMENTATION_CONTRADICTION nenne mindestens ein
kurzes wörtliches Zitat in evidence_spans und setze missing_attributes auf [];
für REQUIRED_ATTRIBUTE_MISSING nenne die fehlenden Angaben in missing_attributes.
DECISION_BASIS_INCOMPLETE ist der technischen und deterministischen
Verarbeitung vorbehalten und darf von dir nicht gewählt werden.
Bewerte weder Höchstmengen noch Ausschlüsse; diese werden
separat deterministisch geprüft. Evidence-Spans müssen kurze, wörtliche und
unveränderte Zitate aus case_documentation sein. Erfinde keine Zitate.
Begründe die Bewertung knapp auf Deutsch.
Gib die Felder in dieser Reihenfolge aus: evidence_spans, missing_attributes,
explanation, zuletzt basis_code.
\end{lstlisting}

\begin{lstlisting}[
  caption={Assessor: user message (example, item 1 of case c02)},
  label={lst:assessor-user-c02-item1},
  basicstyle=\small\ttfamily,
  breaklines=true,
  breakatwhitespace=true,
  columns=fullflexible,
  keepspaces=true,
  showstringspaces=false,
  literate=
    {ä}{{\"a}}1
    {ö}{{\"o}}1
    {ü}{{\"u}}1
    {Ä}{{\"A}}1
    {Ö}{{\"O}}1
    {Ü}{{\"U}}1
    {ß}{{\ss}}1
]
{
  "case_documentation": "Ein Medikament wurde intramuskulär injiziert. Eine venöse Blutentnahme erfolgte ausdrücklich nicht.",
  "claimed_item": {
    "code": "I01",
    "count": 1
  },
  "claimed_code_total": 1,
  "applicable_rule": {
    "code": "I01",
    "legend": "Venöse Blutentnahme.",
    "documentation_requirements": [
      "Probenart (Blut)",
      "Entnahmeweg (venös)"
    ]
  }
}
\end{lstlisting}

\begin{lstlisting}[
  caption={Baseline: system prompt},
  label={lst:baseline-system-prompt},
  basicstyle=\small\ttfamily,
  breaklines=true,
  breakatwhitespace=true,
  columns=fullflexible,
  keepspaces=true,
  showstringspaces=false,
  literate=
    {ä}{{\"a}}1
    {ö}{{\"o}}1
    {ü}{{\"u}}1
    {Ä}{{\"A}}1
    {Ö}{{\"O}}1
    {Ü}{{\"U}}1
    {ß}{{\ss}}1
]
Du prüfst einen Abrechnungsfall vollständig in einem Schritt. Du erhältst die
Falldokumentation, die beanspruchten Positionen mit Menge und den gesamten
Regelkatalog mit allen Releases und ihren Gültigkeitszeiträumen.

Ein Release gilt, wenn valid_from <= effective_date < valid_to;
bei valid_to = null gibt es keine obere Grenze.
Wenn kein oder mehr als ein Release gilt, ist das Gesamtergebnis NOT_EVALUABLE.
Gib dann für jede Position NO_APPLICABLE_RULE_CONTEXT aus.

Bei genau einem gültigen Release prüfe für jede Position, ob ihr Code darin
existiert. Für einen Code ohne Eintrag wähle NO_APPLICABLE_RULE_CONTEXT;
ein Dokumentationsurteil ist für diese Position nicht möglich.
Prüfe für die bekannten Codes Höchstmengen und gegenseitige Ausschlüsse.
Die Höchstmenge gilt je Code für die Summe der Mengen aller Positionen mit
diesem Code. Beurteile für jede Position mit anwendbarer Regel die
Dokumentation und wähle genau ein basis_code:
- DOCUMENTATION_SUPPORT nur, wenn die Dokumentation die Position positiv belegt.
- DOCUMENTATION_CONTRADICTION nur bei einem dokumentierten Widerspruch oder einer
 ausdrücklich unvereinbaren Alternative.
- REQUIRED_ATTRIBUTE_MISSING, wenn eine für die Zuordnung zur Regel nötige
 Angabe fehlt.
Beurteile die Deckung der gesamten beanspruchten Menge je Code durch die
Dokumentation, auch wenn diese Menge auf mehrere Positionen aufgeteilt ist.
Ein dokumentierter Leistungsvorgang darf nicht mehrfach zur Deckung derselben
Codebeanspruchung gezählt werden.
Das bloße Fehlen eines Widerspruchs ist keine positive Evidenz.
Widersprechen sich Angaben in der Dokumentation zu einer geforderten Angabe,
gilt sie als nicht belastbar dokumentiert: Wähle REQUIRED_ATTRIBUTE_MISSING
und nenne die Angabe.

Gib genau ein Gesamtergebnis aus:
- SUPPORTED: alle Positionen belegt und keine Regel verletzt.
- NOT_SUPPORTED: mindestens ein Widerspruch, eine überschrittene Höchstmenge
 oder ein Ausschluss.
- INSUFFICIENT_DOCUMENTATION: keine Verletzung, aber mindestens eine nötige
 Angabe fehlt.
- NOT_EVALUABLE: kein eindeutig gültiges Release für das Datum oder ein Code
 ohne Eintrag im gültigen Release.
Bei genau einem gültigen Release gilt bei mehreren zutreffenden Ergebnissen
folgende Priorität: NOT_SUPPORTED > NOT_EVALUABLE > INSUFFICIENT_DOCUMENTATION > SUPPORTED.

Antworte ausschließlich im vorgegebenen JSON-Schema, mit einem Urteil je
Position in der Reihenfolge der Positionsnummern. Begründe knapp auf Deutsch.
Gib zuerst items aus, je Position mit position, explanation und zuletzt
basis_code. Danach folgen die Gesamterklärung explanation und zuletzt outcome.
\end{lstlisting}

\section{Synthetic catalog and reference cases}\label{secB1}

\setlength{\LTleft}{0pt}
\setlength{\LTright}{\fill}
\setlength{\LTcapwidth}{\linewidth}
\makeatletter
\renewcommand{\LT@makecaption}[3]{%
  \LT@mcol\LT@cols c{\hbox to\z@{\hss\parbox[t]\LTcapwidth{%
    \raggedright #1{#2\ }#3\par\vskip\baselineskip}\hss}}}
\makeatother
\newcommand{\rowsep}{\specialrule{0.2pt}{1pt}{1pt}}
\newlength{\LTtw}

\begingroup
\small
\setlength{\tabcolsep}{3pt}
\setlength{\LTtw}{\dimexpr\linewidth-8\tabcolsep\relax}
\begin{longtable}{@{}>{\raggedright\arraybackslash}p{.08\LTtw}>{\raggedright\arraybackslash}p{.33\LTtw}>{\raggedright\arraybackslash}p{.34\LTtw}>{\raggedright\arraybackslash}p{.13\LTtw}>{\raggedright\arraybackslash}p{.12\LTtw}@{}}
\caption{Synthetic Catalog. The release labelled 2024 is \texttt{GOAE-SYN-2024}, valid from 2024-01-01 up to but excluding 2025-01-01. The release labelled 2025 is \texttt{GOAE-SYN-2025}, valid from 2025-01-01 with no upper bound. All table fields apply to both releases except for the two maximum-count values. The catalog is entirely synthetic and provides no legal basis for billing.}\label{tab:complete_catalog}\\
\toprule
Code & Legend & Documentation requirements & Maximum\newline(2024/2025) & Exclusions \\
\midrule
\endfirsthead
\toprule
Code & Legend & Documentation requirements & Maximum\newline(2024 / 2025) & Exclusions \\
\midrule
\endhead
\midrule
\endfoot
\botrule
\endlastfoot
B01 & Kurze persönliche Beratung. & Kontaktart (persönlich); Beratungsumfang (kurz) & 1 / 1 & B02\newline B03 \\
\rowsep
B02 & Eingehende, das gewöhnliche Maß übersteigende Beratung (Dauer mindestens 10 Minuten). & Beratungsumfang (ausführlich, eingehend); Beratungsinhalt; Dauer & 1 / 1 & B01\newline B04 \\
\rowsep
B03 & Beratung mittels Fernkommunikation. & Kommunikationsweg (Telefon, Video) & 1 / 1 & B01 \\
\rowsep
B04 & Eingehende Beratung bei chronischem Verlauf. & chronischer Verlauf; Beratungsumfang (ausführlich, eingehend) & 1 / 1 & B02 \\
\rowsep
U01 & Symptombezogene Untersuchung. & Symptombezug (Anlass); untersuchte Region & 1 / 1 & U02\newline U03 \\
\rowsep
U02 & Vollständige Untersuchung eines Organsystems. & Organsystem; Untersuchungsumfang (vollständig) & 1 / 1 & U01 \\
\rowsep
U03 & Umfassende körperliche Statusuntersuchung. Der Ganzkörperstatus beinhaltet die Untersuchung der Haut, der sichtbaren Schleimhäute, der Brust- und Bauchorgane, der Stütz- und Bewegungsorgane sowie eine orientierende neurologische Untersuchung. & Haut; sichtbare Schleimhäute; Brust- und Bauchorgane; Stütz- und Bewegungsorgane; orientierende neurologische Untersuchung & 1 / 1 & U01 \\
\rowsep
I01 & Venöse Blutentnahme. & Probenart (Blut); Entnahmeweg (venös) & 2 / 3 & --- \\
\rowsep
I02 & Intramuskuläre Injektion. & Applikationsweg (intramuskulär) & 3 / 3 & --- \\
\rowsep
I03 & Intravenöse Infusion bis zu 30 Minuten Dauer. & Applikationsweg (intravenös); Infusionsdauer (bis 30 Minuten) & 2 / 2 & I04 \\
\rowsep
I04 & Intravenöse Infusion von mehr als 30 Minuten Dauer. & Applikationsweg (intravenös); Infusionsdauer (über 30 Minuten) & 1 / 1 & I03 \\
\rowsep
W01 & Erstversorgung einer kleinen Wunde. & Wundart (klein, oberflächlich, unkompliziert); Versorgungsmaßnahme (Reinigung, Verband) & 2 / 2 & --- \\
\rowsep
S01 & Ultraschalluntersuchung eines Organs. & untersuchtes Organ & 2 / 2 & --- \\
\rowsep
S02 & Zusätzliche Ultraschalluntersuchung eines weiteren Organs. & weiteres untersuchtes Organ (benannt) & 3 / 3 & --- \\
\end{longtable}
\endgroup

\begingroup
\small
\setlength{\tabcolsep}{3pt}
\setlength{\LTtw}{\dimexpr\linewidth-6\tabcolsep\relax}
\begin{longtable}{@{}>{\raggedright\arraybackslash}p{.13\LTtw}>{\raggedright\arraybackslash}p{.49\LTtw}>{\raggedright\arraybackslash}p{.21\LTtw}>{\raggedright\arraybackslash}p{.17\LTtw}@{}}
\caption{All 36 reference cases. Documentation is reproduced verbatim; claims and verdicts retain their original order. Each claim is shown as code $\times$ count followed by its reference verdict. Repeated codes remain separate positions. A dash denotes the absence of a semantic label or an applicable release; it is not an additional verdict. Item verdicts: S = \texttt{DOCUMENTATION\_SUPPORT}; C = \texttt{DOCUMENTATION\_CONTRADICTION}; M = \texttt{REQUIRED\_ATTRIBUTE\_MISSING}. Rule basis: OK = \texttt{RULE\_CHECKS\_PASSED}; MAX = \texttt{MAX\_COUNT\_EXCEEDED}; EXCL = \texttt{EXCLUSION\_CONFLICT}; NO\_RULE = \texttt{NO\_APPLICABLE\_RULE\_CONTEXT}; NO\_REL = \texttt{RULE\_VERSION\_UNRESOLVED}. Case outcomes: SUP = \texttt{SUPPORTED}; NS = \texttt{NOT\_SUPPORTED}; ID = \texttt{INSUFFICIENT\_DOCUMENTATION}; NE = \texttt{NOT\_EVALUABLE}. The last column lists expected rule basis codes followed by the expected case outcome. All values are reference expectations, not observed model outputs.}\label{tab:complete_cases}\\
\toprule
Case / date / release & Documentation (verbatim) & Claims: reference verdicts & Rule basis $\to$ outcome \\
\midrule
\endfirsthead
\toprule
Case / date / release & Documentation (verbatim) & Claims: reference verdicts & Rule basis $\to$ outcome \\
\midrule
\endhead
\midrule
\endfoot
\botrule
\multicolumn{4}{@{}p{\linewidth}@{}}{\footnotesize
\textsuperscript{1}The known-code position retains a semantic reference label; the unknown-code position has none.\newline
\textsuperscript{2}No applicable release, hence no semantic reference labels.\newline
\textsuperscript{3}Both I01 positions carry contradiction labels because the assessor judges each position against the total claimed for its code: two services are claimed, exactly one is documented, and further services are denied. This does not imply that no blood sampling occurred.}\\
\endlastfoot
a01\newline 2024-02-12\newline 2024 & Die Patientin wurde kurz persönlich zu den aktuellen Beschwerden und zum weiteren Vorgehen beraten. Anschließend wurde venöses Blut entnommen. Eine kleine oberflächliche Wunde wurde gereinigt und mit einem einfachen Verband versorgt. & B01 $\times$ 1: S\newline I01 $\times$ 1: S\newline W01 $\times$ 1: S & OK\newline $\to$ SUP \\
\rowsep
a02\newline 2024-05-03\newline 2024 & Es erfolgte eine vollständige Untersuchung des kardiopulmonalen Organsystems. Danach wurde die Leber sonografisch untersucht. Zusätzlich wurde die rechte Niere als weiteres Organ sonografisch beurteilt. & U02 $\times$ 1: S\newline S01 $\times$ 1: S\newline S02 $\times$ 1: S & OK\newline $\to$ SUP \\
\rowsep
a03\newline 2024-08-19\newline 2024 & Der Patient wurde telefonisch zu den aktuellen Beschwerden und den nächsten Schritten beraten. Später wurde eine kleine unkomplizierte Schnittwunde gereinigt und verbunden. & B03 $\times$ 1: S\newline W01 $\times$ 1: S & OK\newline $\to$ SUP \\
\rowsep
a04\newline 2024-10-02\newline 2024 & Bei bekanntem chronischem Verlauf erfolgte eine eingehende persönliche Beratung zum bisherigen Verlauf und zur Therapie. Anschließend wurde ein Medikament intramuskulär injiziert. & B04 $\times$ 1: S\newline I02 $\times$ 1: S & OK\newline $\to$ SUP \\
\rowsep
a05\newline 2024-12-30\newline 2024 & Es fand eine ausführliche persönliche Beratung von etwa 15 Minuten zu Beschwerden, Befunden und Therapieoptionen statt. Danach wurde venöses Blut entnommen. Eine kleine oberflächliche Wunde wurde gereinigt und einfach verbunden. & B02 $\times$ 1: S\newline I01 $\times$ 1: S\newline W01 $\times$ 1: S & OK\newline $\to$ SUP \\
\rowsep
a06\newline 2025-03-18\newline 2025 & Es wurde ein umfassender körperlicher Status erhoben mit Untersuchung der Haut, der sichtbaren Schleimhäute, der Brust- und Bauchorgane und der Stütz- und Bewegungsorgane sowie einer orientierenden neurologischen Untersuchung. Anschließend wurde die Milz sonografisch untersucht. Zusätzlich wurde die linke Niere als weiteres Organ sonografisch beurteilt. & U03 $\times$ 1: S\newline S01 $\times$ 1: S\newline S02 $\times$ 1: S & OK\newline $\to$ SUP \\
\rowsep
a07\newline 2025-06-11\newline 2025 & Die aktuellen Beschwerden und das weitere Vorgehen wurden telefonisch besprochen. Später wurde eine intravenöse Kurzinfusion über etwa 20 Minuten durchgeführt und beendet. & B03 $\times$ 1: S\newline I03 $\times$ 1: S & OK\newline $\to$ SUP \\
\rowsep
a08\newline 2025-09-01\newline 2025 & Zwei kleine oberflächliche Wunden wurden jeweils gereinigt und einfach verbunden. Danach wurden zwei Medikamente an getrennten Stellen intramuskulär injiziert. & W01 $\times$ 2: S\newline I02 $\times$ 2: S & OK\newline $\to$ SUP \\
\rowsep
a09\newline 2025-10-27\newline 2025 & Die Patientin wurde kurz persönlich zu den aktuellen Beschwerden beraten. Anschließend wurde dreimal getrennt venöses Blut entnommen. & B01 $\times$ 1: S\newline I01 $\times$ 3: S & OK\newline $\to$ SUP \\
\rowsep
b01\footnotemark[1]\newline 2024-03-08\newline 2024 & Eine als Leistung X99 bezeichnete Maßnahme wurde dokumentiert. Zusätzlich wurde venöses Blut entnommen. & X99 $\times$ 1: ---\newline I01 $\times$ 1: S & NO\_RULE\newline $\to$ NE \\
\rowsep
b02\footnotemark[1]\newline 2024-12-29\newline 2024 & Es wurde venöses Blut entnommen. Außerdem wurde eine kleine unkomplizierte Wunde gereinigt und verbunden. & I1 $\times$ 1: ---\newline W01 $\times$ 1: S & NO\_RULE\newline $\to$ NE \\
\rowsep
b03\newline 2024-06-14\newline 2024 & Zu Beginn erfolgte eine kurze persönliche Beratung. Im weiteren Verlauf wurde ein ausführliches persönliches Beratungsgespräch von rund 20 Minuten zu Befunden und Therapie geführt. Danach wurde venöses Blut entnommen. & B01 $\times$ 1: S\newline B02 $\times$ 1: S\newline I01 $\times$ 1: S & EXCL\newline $\to$ NS \\
\rowsep
b04\newline 2025-04-22\newline 2025 & Zunächst wurde eine intravenöse Kurzinfusion über 20 Minuten durchgeführt. Später erfolgte eine intravenöse Langzeitinfusion über 90 Minuten. Zusätzlich wurde venöses Blut entnommen. & I03 $\times$ 1: S\newline I04 $\times$ 1: S\newline I01 $\times$ 1: S & EXCL\newline $\to$ NS \\
\rowsep
b05\newline 2024-11-07\newline 2024 & Im dokumentierten Fall wurden drei getrennte venöse Blutentnahmen durchgeführt. Zusätzlich wurde eine kleine oberflächliche Wunde gereinigt und einfach verbunden. & I01 $\times$ 3: S\newline W01 $\times$ 1: S & MAX\newline $\to$ NS \\
\rowsep
b06\newline 2025-08-30\newline 2025 & Drei kleine oberflächliche Wunden wurden jeweils gereinigt und mit einem einfachen Verband versorgt. Weitere Leistungen wurden nicht durchgeführt. & W01 $\times$ 1: S\newline W01 $\times$ 2: S & MAX\newline $\to$ NS \\
\rowsep
b07\footnotemark[2]\newline 2023-11-20\newline --- & Die Patientin wurde kurz persönlich zu den aktuellen Beschwerden beraten. Anschließend wurde venöses Blut entnommen. & B01 $\times$ 1: ---\newline I01 $\times$ 1: --- & NO\_REL\newline $\to$ NE \\
\rowsep
b08\newline 2025-01-01\newline 2025 & Der Patient wurde vormittags und nachmittags jeweils kurz persönlich zu den aktuellen Beschwerden beraten. Dazwischen wurden die Laborwerte telefonisch besprochen. & B01 $\times$ 2: S\newline B03 $\times$ 1: S & MAX, EXCL\newline $\to$ NS \\
\rowsep
c01\newline 2024-01-15\newline 2024 & Ein Medikament wurde intramuskulär injiziert. Eine symptombezogene Untersuchung wurde ausdrücklich nicht durchgeführt. & U01 $\times$ 1: C\newline I02 $\times$ 1: S & OK\newline $\to$ NS \\
\rowsep
c02\newline 2024-04-09\newline 2024 & Ein Medikament wurde intramuskulär injiziert. Eine venöse Blutentnahme erfolgte ausdrücklich nicht. & I01 $\times$ 1: C\newline I02 $\times$ 1: S & OK\newline $\to$ NS \\
\rowsep
c03\newline 2024-07-24\newline 2024 & Eine kleine oberflächliche Wunde wurde gereinigt und verbunden. Eine Ultraschalluntersuchung wurde nicht durchgeführt. & W01 $\times$ 1: S\newline S01 $\times$ 1: C & OK\newline $\to$ NS \\
\rowsep
c04\newline 2024-12-31\newline 2024 & Die Leber wurde sonografisch untersucht. Weitere Organe wurden ausdrücklich nicht sonografisch beurteilt. Danach wurde venöses Blut entnommen. & S01 $\times$ 1: S\newline S02 $\times$ 1: C\newline I01 $\times$ 1: S & OK\newline $\to$ NS \\
\rowsep
c05\newline 2024-12-31\newline 2024 & Es wurde venöses Blut entnommen. Eine Infusion wurde weder begonnen noch durchgeführt. & I04 $\times$ 1: C\newline I01 $\times$ 1: S & OK\newline $\to$ NS \\
\rowsep
c06\newline 2025-02-17\newline 2025 & Wegen einer akuten Schnittverletzung erfolgte eine kurze Beratung; ein chronischer Verlauf lag nicht vor. Die kleine Wunde wurde gereinigt und einfach verbunden. & B04 $\times$ 1: C\newline W01 $\times$ 1: S & OK\newline $\to$ NS \\
\rowsep
c07\newline 2025-05-28\newline 2025 & Zunächst wurde venöses Blut entnommen. Das Medikament wurde anschließend intravenös verabreicht. & I01 $\times$ 1: S\newline I02 $\times$ 1: C & OK\newline $\to$ NS \\
\rowsep
c08\newline 2024-03-19\newline 2024 & Wegen Knieschmerzen rechts erfolgte eine symptombezogene Untersuchung des rechten Kniegelenks. Die Beratung dazu fand persönlich in der Praxis statt; ein telefonischer Kontakt erfolgte nicht. Eine Sonographie wurde durchgeführt. & U01 $\times$ 1: S\newline B03 $\times$ 1: C\newline S01 $\times$ 1: M & OK\newline $\to$ NS \\
\rowsep
c09\footnotemark[3]\newline 2024-09-16\newline 2024 & Es wurde genau einmal venöses Blut entnommen; weitere Entnahmen erfolgten nicht. Eine kleine oberflächliche Wunde wurde gereinigt und verbunden. & I01 $\times$ 1: C\newline I01 $\times$ 1: C\newline W01 $\times$ 1: S & OK\newline $\to$ NS \\
\rowsep
d01\newline 2024-02-28\newline 2024 & Beschwerden und weiteres Vorgehen wurden besprochen. Anschließend wurde venöses Blut entnommen. & B01 $\times$ 1: M\newline I01 $\times$ 1: S & OK\newline $\to$ ID \\
\rowsep
d02\newline 2024-09-12\newline 2024 & Das Abdomen wurde klinisch untersucht. Danach wurde venöses Blut entnommen. & U02 $\times$ 1: M\newline I01 $\times$ 1: S & OK\newline $\to$ ID \\
\rowsep
d03\newline 2024-12-05\newline 2024 & Der Oberbauch wurde sonografisch untersucht und die Leber sicher beurteilt. & S01 $\times$ 1: S\newline S02 $\times$ 1: M & OK\newline $\to$ ID \\
\rowsep
d04\newline 2024-12-30\newline 2024 & Es fand ein Gespräch über eine bekannte Erkrankung statt. Danach wurde venöses Blut entnommen. & B04 $\times$ 1: M\newline I01 $\times$ 1: S & OK\newline $\to$ ID \\
\rowsep
d05\newline 2025-07-07\newline 2025 & Eine intravenöse Infusion wurde durchgeführt. Zusätzlich wurde venöses Blut entnommen. & I03 $\times$ 1: M\newline I01 $\times$ 1: S & OK\newline $\to$ ID \\
\rowsep
d06\newline 2025-11-21\newline 2025 & Eine kleine oberflächliche Wunde wurde gereinigt und verbunden. Außerdem wurde eine Probe gewonnen. Ein Medikament wurde parenteral verabreicht. & W01 $\times$ 1: S\newline I01 $\times$ 1: M\newline I02 $\times$ 1: M & OK\newline $\to$ ID \\
\rowsep
d07\newline 2025-10-10\newline 2025 & Es erfolgte eine gezielte Sonografie der Leber. Als körperlicher Befund wurde ausschließlich der sonografisch erhobene Leberbefund dokumentiert. Anschließend wurde venöses Blut entnommen. & U03 $\times$ 1: M\newline S01 $\times$ 1: S\newline I01 $\times$ 1: S & OK\newline $\to$ ID \\
\rowsep
d08\newline 2025-03-03\newline 2025 & Eine Wunde wurde versorgt. Es fand ein Beratungsgespräch statt. & W01 $\times$ 1: M\newline B02 $\times$ 1: M & OK\newline $\to$ ID \\
\rowsep
d09\newline 2024-07-08\newline 2024 & Eine Infusion wurde angelegt. Eine Untersuchung erfolgte. & I04 $\times$ 1: M\newline U01 $\times$ 1: M & OK\newline $\to$ ID \\
\rowsep
d10\newline 2025-05-19\newline 2025 & Der Patient wurde zu den Laborwerten beraten. Zuvor wurde venöses Blut entnommen. & B03 $\times$ 1: M\newline I01 $\times$ 1: S & OK\newline $\to$ ID \\
\end{longtable}
\endgroup




\end{appendices}


\bibliography{sn-bibliography}

@article{soroush2024,
  author		= "Soroush, A. and Glicksberg, B. S. and Zimlichman, E. and Barash, Y. and 
					Freeman, R. and Charney, A. W. and Nadkarni, G. N. and Klang, E.",
  title			= "Large language models are poor medical coders---benchmarking of 
					medical code querying",
  journal		= "{NEJM AI}",
  volume		= "1",
  number		= "5",
  pages			= "AIdbp2300040",
  year			= "2024",
  doi			= "10.1056/AIdbp2300040"
}

@article{chomutare2025,
  author		= "Chomutare, T. and Svenning, T. O. and Hern{\'a}ndez, M. {\'A}. T. and 
					Ngo, P. D. and Budrionis, A. and Markljung, K. and Hind, L. I. and 
					Torsvik, T. and Mikalsen, K. {\O}. and Babic, A. and Dalianis, H.",
  title			= "Artificial intelligence to improve clinical coding practice in 
					Scandinavia: crossover randomized controlled trial",
  journal		= "J. Med. Internet Res.",
  volume		= "27",
  pages			= "e71904",
  year			= "2025",
  doi			= "10.2196/71904"
}

@article{arzideh2026,
  author		= "Arzideh, K. and Sch{\"a}fer, H. and Idrissi-Yaghir, A. and Eryilmaz, B. and
					Warmer, S. and Hartmann, E. M. and Borys, K. and Schmidt, C. S. and
					Haubold, J. and Umutlu, L. and Forsting, M. and Nensa, F. and
					Hosch, R.",
  title			= "Comparison of proprietary and fine-tuned large language models for
					multi-label classification of billing codes from radiology reports",
  journal		= "Eur. Radiol.",
  volume		= "36",
  number		= "8",
  pages			= "6218--6231",
  year			= "2026",
  doi			= "10.1007/s00330-026-12445-3"
}

@article{prenosil2025,
  author		= "Prenosil, G. A. and Weitzel, T. K. and Bello, S. C. and Mingels, C. and
					Manzini, G. and Meier, L. P. and Shi, K.-Y. and Rominger, A. and
					Afshar-Oromieh, A.",
  title			= "Neuro-symbolic {AI} for auditable cognitive information extraction
					from medical reports",
  journal		= "Commun. Med.",
  volume		= "5",
  number		= "1",
  pages			= "491",
  year			= "2025",
  doi			= "10.1038/s43856-025-01194-x"
}

@article{amugongo2025,
  author		= "Amugongo, L. M. and Mascheroni, P. and Brooks, S. and Doering, S. and
					Seidel, J.",
  title			= "Retrieval augmented generation for large language models in healthcare:
					a systematic review",
  journal		= "{PLOS} Digit. Health",
  volume		= "4",
  number		= "6",
  pages			= "1--33",
  year			= "2025",
  doi			= "10.1371/journal.pdig.0000877"
}

@inproceedings{gan2025,
  author		= "Gan, Y. and Rybinski, M. and Hachey, B. and Kummerfeld, J. K.",
  title			= "Aligning {AI} research with the needs of clinical coding workflows: 
					eight recommendations based on {US} data analysis and critical review",
  editor		= "Che, W. and Nabende, J. and Shutova, E. and Pilehvar, M. T.",
  booktitle		= "Proceedings of the 63rd Annual Meeting of the Association for 
					Computational Linguistics (Volume 1: Long Papers)",
  pages			= "909--922",
  address		= "Vienna, Austria",
  publisher		= "Association for Computational Linguistics",
  year			= "2025",
  doi			= "10.18653/v1/2025.acl-long.45"
}

@inproceedings{yuan2025,
  author		= "Yuan, M. and Shing, H.-C. and Strong, M. and Shivade, C.",
  title			= "Toward reliable clinical coding with language models: verification and
					lightweight adaptation",
  editor		= "Potdar, S. and Rojas-Barahona, L. and Montella, S.",
  booktitle		= "Proceedings of the 2025 Conference on Empirical Methods in Natural
					Language Processing: Industry Track",
  pages			= "173--184",
  address		= "Suzhou, China",
  publisher		= "Association for Computational Linguistics",
  year			= "2025",
  doi			= "10.18653/v1/2025.emnlp-industry.12"
}

@inproceedings{wang2026,
  author		= "Wang, B. and Shangguan, Z. and Tegtmeyer, K. and Zhang, Z. and
					Chheang, S. and Cohan, A.",
  title			= "{CPTC}oder: a reliable {LLM} system for medical procedure code
					prediction",
  editor		= "Durrett, G. and Jian, P.",
  booktitle		= "Proceedings of the 64th Annual Meeting of the Association for
					Computational Linguistics (Volume 3: System Demonstrations)",
  pages			= "605--614",
  address		= "San Diego, California, United States",
  publisher		= "Association for Computational Linguistics",
  year			= "2026",
  doi			= "10.18653/v1/2026.acl-demo.60"
}

@inproceedings{ge2026,
  author		= "Ge, C. and Zhang, R. and Wang, Y. and Liu, S. and Jiaen, L. and
					Weihuang and Chen, W.",
  title			= "{GuideTree}: guideline-induced review trees for long medical records",
  editor		= "Li, Y. and Rehm, G. and Tu, M.",
  booktitle		= "Proceedings of the 64th Annual Meeting of the Association for
					Computational Linguistics (Volume 6: Industry Track)",
  pages			= "1589--1604",
  address		= "San Diego, California, USA",
  publisher		= "Association for Computational Linguistics",
  year			= "2026",
  doi			= "10.18653/v1/2026.acl-industry.110"
}

@inproceedings{thorne2018,
  author		= "Thorne, J. and Vlachos, A. and Christodoulopoulos, C. and Mittal, A.",
  title			= "{FEVER}: a large-scale dataset for fact extraction and verification",
  editor		= "Walker, M. and Ji, H. and Stent, A.",
  booktitle		= "Proceedings of the 2018 Conference of the North American Chapter of
					the Association for Computational Linguistics: Human Language
					Technologies, Volume 1 (Long Papers)",
  pages			= "809--819",
  address		= "New Orleans, Louisiana",
  publisher		= "Association for Computational Linguistics",
  year			= "2018",
  doi			= "10.18653/v1/N18-1074"
}

@misc{shirey2007,
  author		= "Shirey, R.",
  title			= "Internet security glossary, version 2",
  howpublished		= "{RFC} 4949",
  publisher		= "{RFC} Editor",
  year			= "2007",
  doi			= "10.17487/RFC4949"
}

@misc{coalson2026,
  author		= "Coalson, Z. and Sohler, B. and Gabriel, A. and Hong, S.",
  title			= "Fail-closed alignment for large language models",
  year			= "2026",
  note			= "Preprint at \url{https://arxiv.org/abs/2602.16977v1}"
}

\end{document}